\documentclass[aps,prl,reprint,groupedaddress,showpacs]{revtex4-1}
\usepackage{graphicx}% Include figure files
\usepackage{dcolumn}% Align table columns on decimal point
\usepackage{bm}% bold math
\usepackage[latin1]{inputenc}
\begin{document}

%\preprint{}

\title{Observation of nonspecular effects for Gaussian Schell-model light beams}


\author{Michele Merano}
\email[]{michele.merano@unipd.it}
\author{Gabriele Umbriaco}
\author{Giampaolo Mistura}

%\thanks{}
\affiliation{Dipartimento di Fisica e Astronomia G. Galilei, Università degli studi di Padova, via Marzolo 8, 35151 Padova, Italy}


\date{\today}

\begin{abstract}
We investigate experimentally the role of spatial coherence on optical beam shifts. This topic has been the subject of recent theoretical debate. We consider Gaussian Schell-model beams, with different spatial degrees of coherence, reflected at an air-glass interface. We prove that the angular Goos-Hänchen and the angular Imbert-Fedorov effects are affected by the spatial degree of coherence of the incident beam, whereas the spatial Goos-Hänchen effect does not depend on incoherence. Our data unambiguously resolve the theoretical debate in favour of one specific theory.
\end{abstract}

%\pacs{42.25.Kb, 42.25.Gy, 42.30.Ms}
%\keywords{}

\maketitle

\section{}
When a beam of light is reflected from a planar interface it may suffer spatial or angular deviations with respect to the predictions of geometrical optics. These deviations amount to either the Goos-Hänchen (GH) \cite{Goos47} or the Imbert-Fedorov (IF) \cite{Imbert72, Fedorov55} shift depending on whether they occur in either the plane of incidence or in a direction perpendicular to it. The theoretical and experimental investigation of these nonspecular effects has witnessed outstanding results in recent years. Among them there are the first experimental observation of angular shifts in optics \cite{Merano09}, and the theoretical connection in between the IF effect and the spin-Hall effect of light \cite{Onoda04, Bliokh06} that was also experimentally confirmed \cite{Kwiat08}. Studies have not been limited to Gaussian beams, but have been extended to Laguerre-Gaussian beams too \cite{Merano10}. These effects can be observed also in structured interfaces like gratings \cite{Bretenaker01}, photonic crystals \cite{Felbacq04}, waveguides \cite{Boardman07} or resonators \cite{Martina06}. In view of possible applications, beam shifts are exploited for sensing \cite{Hesselink06}. Non specular phenomena are not limited to light waves but have been predicted and observed for matter waves of neutrons \cite{Haan10} and electrons\cite{Bliokh12}.

 Although all the experimental work in this field has been performed till now with light beams that are spatially coherent, the role of incoherence in optical beam shifts has been the subject of recent theoretical debate \cite{Wang12, Aiello12}. Theories have addressed this topic by considering the specific case of Gaussian Schell-model (GS) beams that allows for an analytic treatment. Simon et al.\cite{Tamir89} study beam reflection on multilayer structures and find that the angular GH effect depends on the degree of coherence of the incident light while the spatial GH does not. Wang et al. \cite{Zubairy08} consider the case of total internal reflection (TIR) and claim that the spatial GH shift depends on incoherence. Aiello et al. \cite{Aiello11} present a general theory, valid for any interface, considering spatial and angular GH and IF effects. They find that only the angular part of both the GH and IF shifts is affected by the spatial coherence of the beam.

In this paper we address the role of spatial coherence in beam shifts experimentally. We study the nonspecular effects of GS beams at an air-glass interface. Gaussian Schell-model beams are particularly useful from an experimental point of view because in spite of their partial spatial coherence they are highly directional \cite{Gori78}. This property makes them very useful in studying nonspecular effects. More specifically, we investigate the spatial GH effect and the angular GH and IF effects, and we compare our experimental results with the existing theories. We find that our experimental data agree well with theory in refs. \cite{Tamir89, Aiello11}.

We resume briefly the relevant aspects of the theories that we will address in our experiment. A monochromatic GS beam is a beam of light generated by a planar source whose intensity distribution $I(\textbf{r})$ and spatial degree of coherence $g(\textbf{r})$ are both gaussian:
\begin{equation}
I(\textbf{r})=A\exp\left(-\frac{r^{2}}{2\sigma_{s}^{2}}\right),\quad g(\textbf{r})=\exp\left(-\frac{r^{2}}{2\sigma_{g}^{2}}\right)
\end{equation}
Here $\textbf{r}$ is the position vector in the source plane, $A$ is the intensity amplitude, $\sigma_{s}$ is the beam radius and $\sigma_{g}$ is the correlation length. When the GS beam propagates along the axis \emph{z}, these quantities evolves as $\sigma_{s}(z)$ and $\sigma_{g}(z)$. An important result for GS beams is that the ratio $\sigma_{s}(z)/\sigma_{g}(z)$  is a constant in propagation \cite{Mandel}. In ref. \cite{Zubairy08} authors claim that the spatial GH shift decreases along with the ratio  $\sigma_{g}/ \sigma_{s}$. So the effect predicted is not dependent on beam propagation but only on intrinsic properties of a GS beam. As already mentioned this result is in contrast with refs. \cite{Tamir89, Aiello11} where no effect of incoherence on spatial shifts is predicted, and where angular shifts are also considered. The theory of Aiello et al. in particular shows that the angular shifts for an incoherent beam are expressible in terms of the corresponding shifts calculated for a coherent beam \cite{Aiello08}. The only difference in the expressions for the shifts is the following: while the angular shifts of a Gaussian beam scale with $\theta_{0}^{2}/2$ where $\theta_{0}$ is the angular spread of the Gaussian beam, the angular shifts of a GS beam scale with $\theta_{s}^{2}$:
\begin{equation}
\theta_{s}^{2}=\frac{2}{k^{2}}\left[\frac{1}{(2\sigma_{s}^{2})}+\frac{1}{\sigma_{g}^{2}}\right]
\end{equation}
where $\theta_{s}$ is the angular spread of the GS beam and $k=2\pi / \lambda$ ($\lambda$ is the wavelength of the light).

Our experimental set up is sketched in Fig. 1. A GS beam is incident at a given angle on a right angle BK7 prism. The spatial GH effect is observed in TIR. The angular GH and IF effects are observed in external (air-glass) reflection. All these phenomena are polarization dependent. We observe them by varying the polarization of the incident beam and by measuring the beam displacement after reflection with a position sensitive detector (PSD).

\begin{figure}
\includegraphics{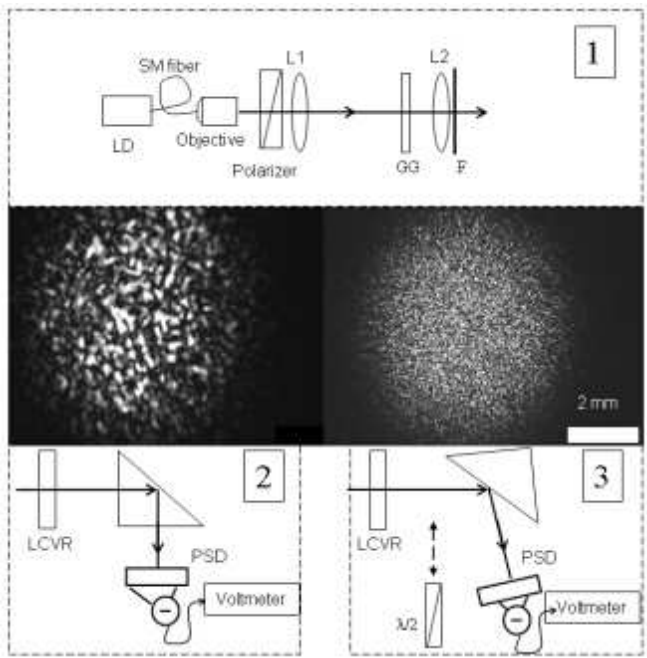}
\caption{\label{}\emph{Panel 1}. GS beam generation. As a monochromatic light source we use a laser diode (LD). A single mode (SM) fiber and a collimating objective generate a high quality Gaussian beam. The polarization is fixed to \emph{p} with a polarizer. A lens (L1) is used to focus the light to a desired spot size on the ground glass (GG). The light scattered by the GG is collimated by the lens L2 covered by an amplitude filter (F). \emph{Inset}. Images of two GS beams that we used for the experiment (open squares and open triangles in Fig.2). They were obtained with a CCD camera (not shown) placed close to F. We verified experimentally that the GS beams evolve with the propagation distance as expected. \emph{Panel 2}. Measurements for the spatial GH shift are taken in TIR configuration. A Liquid crystal variable retarder (LCVR) is used to switch the polarization from \emph{p} to \emph{s}. A position sensitive detector (PSD) measures the small beam displacements. \emph{Panel 3} Measurements for the angular GH shift are taken in external reflection. For measuring the angular IF effect a half wave plate ($\lambda/2$) is inserted.}
\end{figure}

Our light source is a single mode fiber-pigtailed laser at a wavelength of 635 nm, collimated with a 20x objective ($1/e^{2}$ beam waist = 675 $\mu$m) and focussed with a lens L1 (f = 40 cm) on a ground glass (220 grit polishes). The incoherent light so generated is then collimated with a lens L2 (focal length 10 cm). The intensity width of the GS beam is fixed by home-made gaussian amplitude filters (obtained as photographic replica of a computer generated gaussian amplitude profiles) placed in closed proximity to L2. With a CCD camera we verified that the beam radius of our GS beam evolve with the propagation distance as expected \cite{Mandel}. This provides a measurement of $\theta_{s}$ and as as a consequence of $\sigma_{g}$ since for our beams $\sigma_{s} \gg \sigma_{g}$. With this approach we have produced the GS beams with $\sigma_{g}$ = 155 $\mu$m. (one is shown on the left hand side of the inset of fig1.) Our interest is limited to spatial coherence and not to temporal coherence so we do not need to rotate (as required by other experiments)  the ground glass in order to generate a pseudo thermal source and to achieve control of the temporal coherence \cite{Martienssen64}. As a more stringent text, we also tried to remove the lens L1 and to illuminate two diffrent ground glasses (220 and 600 grit polishes) with the collimated beam. For a better collection efficiency we used lenses L2 of focal length 5 cm and 3 cm respectively. With this approach we have produced GS beams with an order of magnitude smaller $\sigma_{g}$ (right hand side of the inset of fig1).

%The GS beam is produced starting from a spatially incoherent source and using a collimating lens (L2, focal length 10 cm and 5 cm) and an amplitude filter that covers it. The source is a ground glass illuminated by a Gaussian beam generated by a single mode fiber-pigtailed laser diode at 635 nm. An objective collimates the Gaussian beam exiting from the fiber. A lens (of variable focal) focuses the beam on the ground glass to a desired spot size. The beam radius of the GS beam is fixed by the amplitude filter. This was obtained as a photographic replica of a computer generated gaussian amplitude profile.

\begin{figure}
\includegraphics{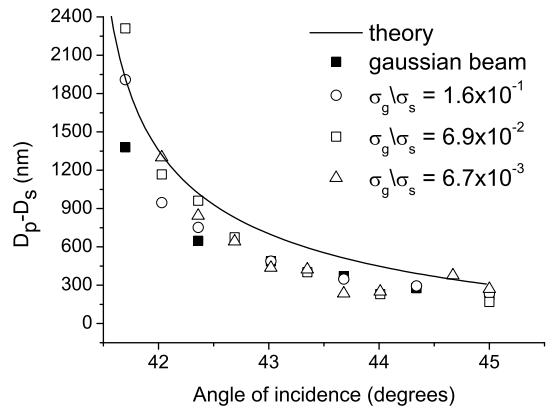}
\caption{\label{} Comparison of the spatial GH effect for a Gaussian beam and three GS beams. Solid squares are the data for the Gaussian beam. The solid line is the theoretical prediction for a Gaussian beam. The other points are for three GS beams with different $\sigma_{g}/ \sigma_{s}$ ratio.  Within our experimental precision we do not see a decrease of the GH signal along with the $\sigma_{g}/ \sigma_{s}$ ratio.}
\end{figure}

The polarization of the light beam exiting the laser diode is fixed to \emph{p} with a polarizer. A liquid crystal variable retarder (LCVR) is used to switch the polarization of the incident light from \emph{p} to \emph{s}. The LCVR is placed immediately after the amplitude filter. It operates at a frequency of 1 Hz. For measuring the angular IF a  $\lambda/2$ is inserted after the LCVR. The GS beam is then incident on a right angle BK7 (n=1.515 at 635 nm) prism and the reflected light is collected by the PSD in combination with a lock-in amplifier (for small displacement measurements) or a voltmeter.

In Fig 2. we report our experimental data for the spatial GH shift. The horizontal axis is the angle of incidence, the vertical axis is the GH shift of a \emph{p} polarized beam with respect to a \emph{s} polarized beam. We compare the experimental results obtained for a Gaussian beam with those obtained for three GS beams with different  $\sigma_{g}/ \sigma_{s}$ ratios. The line in the figure represents the theoretical predictions for a gaussian beam \cite{Aiello08} to be compared with the corresponding experimental data (solid squares). If we focus now on the ensemble of our experimental data, we do not find that the spatial GH shift for GS beams decreases along with the ratio  $\sigma_{g}/ \sigma_{s}$. The correlation lengths and the beam radii for our three GS beams are the following: $\sigma_{g}$ = 155 $\mu$m and $\sigma_{s}$ = 0.9 mm (open circles beam), $\sigma_{g}$ = 155 $\mu$m and $\sigma_{s}$ = 2.2 mm (open squares beam) and $\sigma_{g}$ =15 $\mu$m an $\sigma_{s}$ =2.2 mm (open triangles beam). For these measurements we avoided deliberately angles of incidence so close to the critical angle that part of the beam is transmitted, because we limit our study to TIR.

\begin{figure}
\includegraphics{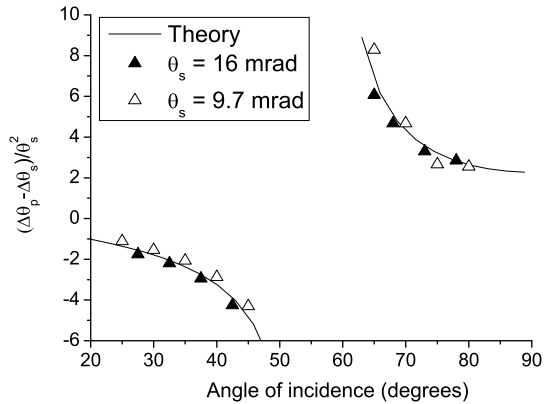}
\caption{\label{} Angular GH shift for two GS beams. Data (solid and open triangles) are compared with theory (solid line) from ref. \cite{Aiello11}.}
\end{figure}

Now we address the angular shifts. In figure 3 we show our data for the angular GH shift as a function of the angle of incidence. Our data refer to two GS beams with different $\theta_{s}$. White triangles data are for the same GS beam of Fig.2. Black triangles data are for a GS beam with  $\sigma_{g}$ =9 $\mu$m an $\sigma_{s}$ =2.2 mm. On the vertical axis we report the angular shift of a \emph{p} polarized beam with respect to a \emph{s} polarized beam. In our experimental approach the angular shift is defined as the beam spatial deviation measured by the PSD, divided by the distance of the PSD from the amplitude filter. To verify the prediction that the angular GH shift scale with $\theta_{s}^{2}$, data have been made dimensionless by dividing them by  $\theta_{s}^{2}$ and by checking that they lie on the same curve. Theoretical prediction are obtained from ref. \cite{Aiello08} according to the rule $\theta_{0}^{2}/2 \rightarrow \theta_{s}^{2}$ mentioned before and predicted in ref. \cite{Aiello11}. The agreement in between data and theory is excellent.

For measuring the angular IF shift it is convenient to switch the linear polarization of the incident beam from -45\textdegree to +45\textdegree because in this case the shift has opposite sign. For this measurement we have only one set of data only but they are good enough to confirm that even for the angular IF effect the agreement in between experiment and theory is good. In the vertical axis of Fig. 4 we do not report the angular shift but our rough data i.e. the beam displacement of a -45\textdegree linear polarized GS beam with respect to a + 45\textdegree linear polarized one at a distance of 27 cm from the amplitude filter. The theoretical solid line is computed from formulas in ref. \cite{Aiello08} by applying again the $\theta_{0}^{2}/2 \rightarrow \theta_{s}^{2}$ rule of ref \cite{Aiello11}. We do not address at all the case of the spatial IF shift in this paper because it is too small to be detected with our present spatial resolution.

\begin{figure}
\includegraphics{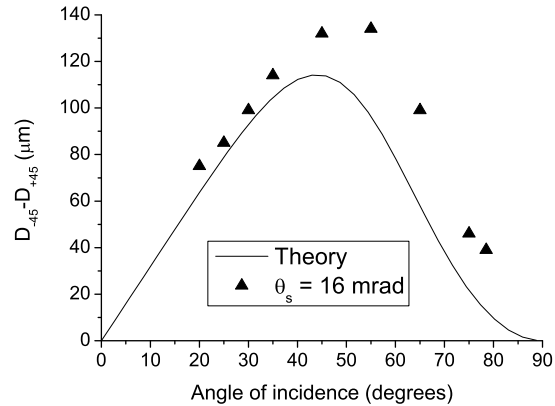}
\caption{\label{} Angular IF shift for a GS beam. Data (solid triangles) are compared with theory (solid line) from ref. \cite{Aiello11}.}
\end{figure}

In conclusion we have presented our experimental data for the GH and the IF shifts for GS beams. For what concerns the spatial GH shift, within our experimental errors, we observe the same effect for spatially coherent and partial spatially coherent beams. Our results resolve the recent theoretical controversy \cite{Wang12, Aiello12} on the spatial GH shift in favor of the theories exposed in refs \cite{Tamir89, Aiello11}. For what concerns the angular GH and IF shifts we show that these last ones are affected by the presence of incoherence. We confirm that these shifts scale with the angular spread of the Gaussian Schell-model beam \cite{Aiello11}.

When finalizing the present work, it appeared on the arXiv a work from L\"{o}ffler et al. on the same subject: arXiv:1207.4364v1.

\begin{acknowledgments}
We thank Giorgio Delfitto for his technical assistance. This work has been partially supported MIUR-PRIN contract 2008Y2P573 and PRAT-UNIPD contract CPDA 118428
\end{acknowledgments}

\bibliography{letter}

\end{document}